\documentclass[preprint]{aastex61}
\usepackage{multirow}
\usepackage{hhline}

\submitjournal{ApJ}

\shorttitle{M80 LE-MP radial distributions}
\shortauthors{Dalessandro et al.}
\begin{document}

\title{The Peculiar Radial Distribution of Multiple Populations in the massive globular cluster M~80}

\correspondingauthor{Emanuele Dalessandro}
\email{emanuele.dalessandro@oabo.inaf.it}

%\author[0000-0002-0786-7307]{Greg J. Schwarz}
%\affil{INAF-OABO\\
%Via Gobetti 93/3 - 40129
%Bologna - Italy
%}

\author{E. Dalessandro}
\affiliation{INAF - Astrophysics and Space Science Observatory Bologna, Via Gobetti 93/3 40129 Bologna - Italy}
%\collaboration{(AAS Journals Data Scientists collaboration)}

\author{M. Cadelano}
\affiliation{Dipartimento di Fisica e Astronomia,  Via Gobetti 93/2 40129 Bologna - Italy}
\affiliation{INAF - Astrophysics and Space Science Observatory Bologna, Via Gobetti 93/3 40129 Bologna - Italy}
%\nocollaboration

\author{E. Vesperini}
\affiliation{Department of Astronomy, Indiana University, Swain West, 727 E. 3rd Street, IN 47405 Bloomington - USA}
%\affiliation{American Astronomical Society \\
%2000 Florida Ave., NW, Suite 300 \\
%Washington, DC 20009-1231, USA}

\author{M. Salaris}
\affiliation{Astrophysics Research Institute, Liverpool John Moores University, IC2 Liverpool Sceince Park, 146 Brownlow Hill, L3 5RF, Liverpool}

\author{F. R. Ferraro}
\affiliation{Dipartimento di Fisica e Astronomia,  Via Gobetti 93/2 40129 Bologna - Italy}
\affiliation{INAF - Astrophysics and Space Science Observatory Bologna, Via Gobetti 93/3 40129 Bologna - Italy}
%\affiliation{TeXnology Inc.}
%\collaboration{(LaTeX collaboration)}

\author{B. Lanzoni}
\affiliation{Dipartimento di Fisica e Astronomia,  Via Gobetti 93/2 40129 Bologna - Italy}
\affiliation{INAF - Astrophysics and Space Science Observatory Bologna, Via Gobetti 93/3 40129 Bologna - Italy}

\author{S. Raso}
\affiliation{Dipartimento di Fisica e Astronomia,  Via Gobetti 93/2 40129 Bologna - Italy}
\affiliation{INAF - Astrophysics and Space Science Observatory Bologna, Via Gobetti 93/3 40129 Bologna - Italy}
%\affiliation{IOP Senior Publisher for the AAS Journals}
%\affiliation{IOP Publishing, Washington, DC 20005}

\author{J. Hong}
\affiliation{Department of Astronomy, Indiana University, Swain West, 727 E. 3rd Street, IN 47405 Bloomington - USA}
\affiliation{Kavli Institute for Astronomy and Astrophysics, Peking University, Yi He Yuan Lu 5, HaiDian District, Beijing 100871, China}

\author{J. J. Webb}
\affiliation{Department of Astronomy, Indiana University, Swain West, 727 E. 3rd Street, IN 47405 Bloomington - USA}
\affiliation{Department of Astronomy \& Astrophysics, University of Toronto, 50 St. George Street, Toronto, ON M5S 3H4, Canada}

\author{A. Zocchi}
\affiliation{Dipartimento di Fisica e Astronomia,  Via Gobetti 93/2 40129 Bologna - Italy}
\affiliation{INAF - Astrophysics and Space Science Observatory Bologna, Via Gobetti 93/3 40129 Bologna - Italy}
\affiliation{European Space Research and Technology Centre, Keplerlaan 1, 2200 AG Noordwijk, Netherlands}

%% Note that the \and command from previous versions of AASTeX is now
%% depreciated in this version as it is no longer necessary. AASTeX 
%% automatically takes care of all commas and "and"s between authors names.

%% AASTeX 6.1 has the new \collaboration and \nocollaboration commands to
%% provide the collaboration status of a group of authors. These commands 
%% can be used either before or after the list of corresponding authors. The
%% argument for \collaboration is the collaboration identifier. Authors are
%% encouraged to surround collaboration identifiers with ()s. The 
%% \nocollaboration command takes no argument and exists to indicate that
%% the nearby authors are not part of surrounding collaborations.

%% Mark off the abstract in the ``abstract'' environment. 
\begin{abstract}

We present a detailed analysis of the radial distribution of light-element multiple populations (LE-MPs)
in the massive and dense globular cluster M~80 based on the combination of UV and optical 
{\it Hubble Space Telescope} data. 
Surprisingly, we find that first generation stars (FG) are significantly more centrally concentrated than extreme second generation ones (SG) out 
to $\sim 2.5 r_h$ from the cluster center. 
To understand the origin of such a peculiar behavior, we used a set of N-body simulations
following the long-term dynamical evolution of LE-MPs.
We find that, given the advanced dynamical state of the cluster, the observed difference does not depend on 
the primordial relative distributions of FG and SG stars.
On the contrary, a difference of $\sim 0.05-0.10 M_{\odot}$ between the average masses of the two sub-populations is needed
to account for the observed radial distributions. We argue that such a mass difference might be the result of the higher He abundance 
of SG stars (of the order of $\Delta Y\sim 0.05-0.06$) with respect to FG.
Interestingly, we find that a similar He variation is necessary to reproduce the horizontal branch morphology of M~80.
These results demonstrate that differences in mass among LE-MPs, due to different He content, 
should be properly taken into account for a correct interpretation of their radial distribution, 
at least in dynamically evolved systems.

\end{abstract}

%% Keywords should appear after the \end{abstract} command. 
%% See the online documentation for the full list of available subject
%% keywords and the rules for their use.

%% From the front matter, we move on to the body of the paper.
%% Sections are demarcated by \section and \subsection, respectively.
%% Observe the use of the LaTeX \label
%% command after the \subsection to give a symbolic KEY to the
%% subsection for cross-referencing in a \ref command.
%% You can use LaTeX's \ref and \label commands to keep track of
%% cross-references to sections, equations, tables, and figures.
%% That way, if you change the order of any elements, LaTeX will
%% automatically renumber them.

%% We recommend that authors also use the natbib \citep
%% and \citet commands to identify citations.  The citations are
%% tied to the reference list via symbolic KEYs. The KEY corresponds
%% to the KEY in the \bibitem in the reference list below. 

\section{Introduction} \label{sec:intro}

Almost all massive ($> 4-5 \times 10^4 M_{\odot}$) and relatively old ($> 2$ Gyr)
globular clusters (GCs) studied with spectroscopic or photometric observational investigations 
have been shown to host light-element multiple populations (LE-MPs) characterized by different abundances in a number 
of light-elements (e.g. C, N, Na, O, He, Al, Mg). 

LE-MPs appear to be ubiquitous as they have been directly observed 
in Galactic globular clusters \citep[e.g.][for a review]{gratton12} as well as 
in external systems, such as the Magellanic Clouds and the Fornax dwarf galaxy \citep{mucciarelli08,larsen14,dalessandro16}. 
Moreover, their presence has been indirectly constrained 
in the GC systems of M31 and M87 \citep{chung11,schiavon13}. 

LE-MPs manifest themselves in very different ways in the color-magnitude diagrams (CMDs) when appropriate 
filter combinations are used. In particular, (near-)UV filters are
efficient in separating LE-MPs in CMDs as variations of the
OH, CN, CH molecular bands have particularly strong effects in the spectral range $3000<\lambda \, (\AA)<4000$ \citep{sbordone11}. 
Indeed, recent and extensive UV Hubble Space Telescope observations, like 
{\it The UV Legacy Survey of Galactic Globular Clusters} \citep{piotto15}, have allowed a significant leap
in our understanding of LE-MPs and their census in GCs. 
These observations have shown that GCs can host from two up to seven (photometrically distinct) stellar populations and their relative number ratios can vary from one
cluster to another with some dependence on cluster mass \citep{piotto15,milone17}.

CN-weak, Na-poor stars are commonly referred to us as first generation/population (FG) and the CN-strong, 
Na-rich ones as second generation/population (SG). Both are believed to have formed during the very early epoch of GC formation and evolution ($\sim 10-100$ Myr -- \citealt{decressin07,dercole08,demink09,denissenkov14}).
However, no consensus has been reached on the multiple population formation history yet and we still lack a complete self-consistent explanation of the physical processes at the basis of the LE-MP and GC formation.

The key to shedding new light on the MP phenomenon is a comprehensive description of their 
properties by means of state-of-the-art photometry and spectroscopy combined with an in depth characterization of their kinematics.
Indeed a number of theoretical studies have predicted that SG and FG stars would form with different initial spatial and kinematical properties (see e.g. \citealt{dercole08}) and some clusters could still retain some memory of these primordial differences (see e.g. \citealt{mastrobuono12,vesperini13,henault15}).

Observational evidence of kinematic differences between FG and SG stars have been recently found in a few 
clusters 
(see \citealt{richer13,bellini15,cordero17}) based on proper motions and radial velocities.
In addition, SG stars are generally observed to be more centrally concentrated than FG ones \citep[e.g.][for recent results]{bellini15,dalessandro16,massari16}, while only in a few clusters, namely NGC~6362 \citep{dalessandro14}, NGC~6121, NGC~6752 \citep{nardiello15} and M~13 \citep{savino18},
FG and SG stars show the same radial distribution. Such a lack of difference in the radial distributions is typically interpreted 
as the result of a cluster's advanced dynamical evolution 
\citet[][see also \citealt{miholics15}]{vesperini13} and significant mass-loss due to both two-body relaxation and interaction with the host galaxy.

In this context, very peculiar is the case of M~15 \citep{larsen15} where
FG stars are found to be more centrally concentrated than the SG out to the cluster's half-light radius, 
while beyond this radius the trend is inverted and the SG population is more concentrated than the FG one \citep{lardo11}. However, this result has been recently questioned by Nardiello et al. (2018).

Following on the variety of results revealed by these observational studies and the possible constraints they provide for our understanding of GC formation and dynamical evolution, in this paper we report on the radial distribution of LE-MPs in the massive and dense GC NGC~6093 (M~80). 
The presence and classification of LE-MPs in M~80 have been previously discussed
by \citet{piotto15} and \citet{milone17} and we refer the readers to these papers for further details.
Based on both the number and radial distribution of its Blue Straggler Star (BSS) population, \citet{ferraro12} classified M~80
as a {\it dynamical old} (i.e. dynamically evolved) stellar system (see also \citealt{lanzoni16}). Indeed it has been suggested 
to be in a transient dynamical state during which stellar interactions are delaying the core-collapse process leading to the 
observed very high fraction of BSS \citep{ferraro99}. 

The paper is structured as follows. In Section~2 the adopted data-sets and data-reduction procedures are described.
Section~3 reports on the LE-MP selection in the CMD and Section~4 on their radial distributions.
In Section~5 we present the results of a set of N-body models aimed at providing some theoretical guidance on the interpretation of our observational results. In Section~6 we analyze the horizontal branch morphology of the cluster to constrain the maximum He variation among LE-MPs. In Section~7 we summarize the main results.

\begin{figure}
\plotone{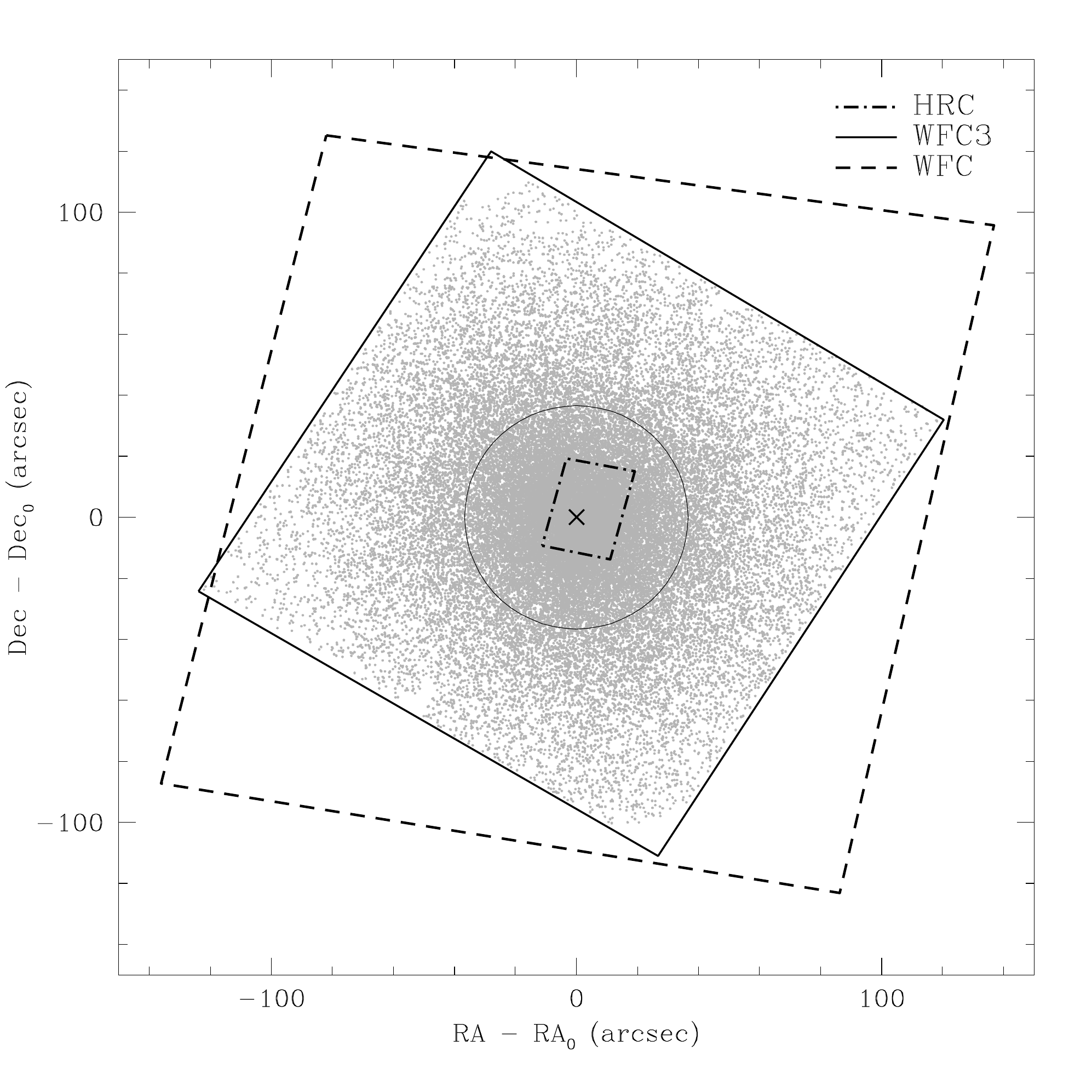}
\caption{Map of the HST database used for the radial distribution analysis of the LE-MPs respect to the position of $C_{\rm grav}$ indicated with the cross.}
\end{figure} 

\section{Observations and data analysis}

To study the radial distribution of M~80 we used a combination of HST WFC3/UVIS images 
acquired in the F275W, F336W and F438W bands through proposal GO-12605
(PI: Piotto), data obtained with the ACS/WFC in the F606W and F814W bands 
(GO-10775, PI: Sarajedini) and
one ACS/HRC image obtained in the F250W band acquired 
through proposal GO-10183 (PI: Knigge). 
Details about each data-set are reported in Table~1.
Appropriate dither
patterns of tens of arcseconds have been adopted for each pointing of the WFC3/UVIS and ACS/WFC data-sets in order to
fill the inter chip gaps and avoid spurious effects due to bad pixels.

For WFC3/UVIS and ACS/WFC samples we used images processed, flat-fielded,
bias subtracted and corrected for charge transfer efficiency ({\tt CTE}) by the
standard HST pipeline ({\tt flc} images). The most updated pixel-area-maps ({\tt PAM}
images) were applied independently to each chip and image. 
For the ACS/HRC photometry, we used the flat-fielded, 
bias subtracted, {\tt CTE} corrected and cosmic ray rejected image produced by the standard
HST pipeline ({\tt crj} image).

\begin{figure}
\plotone{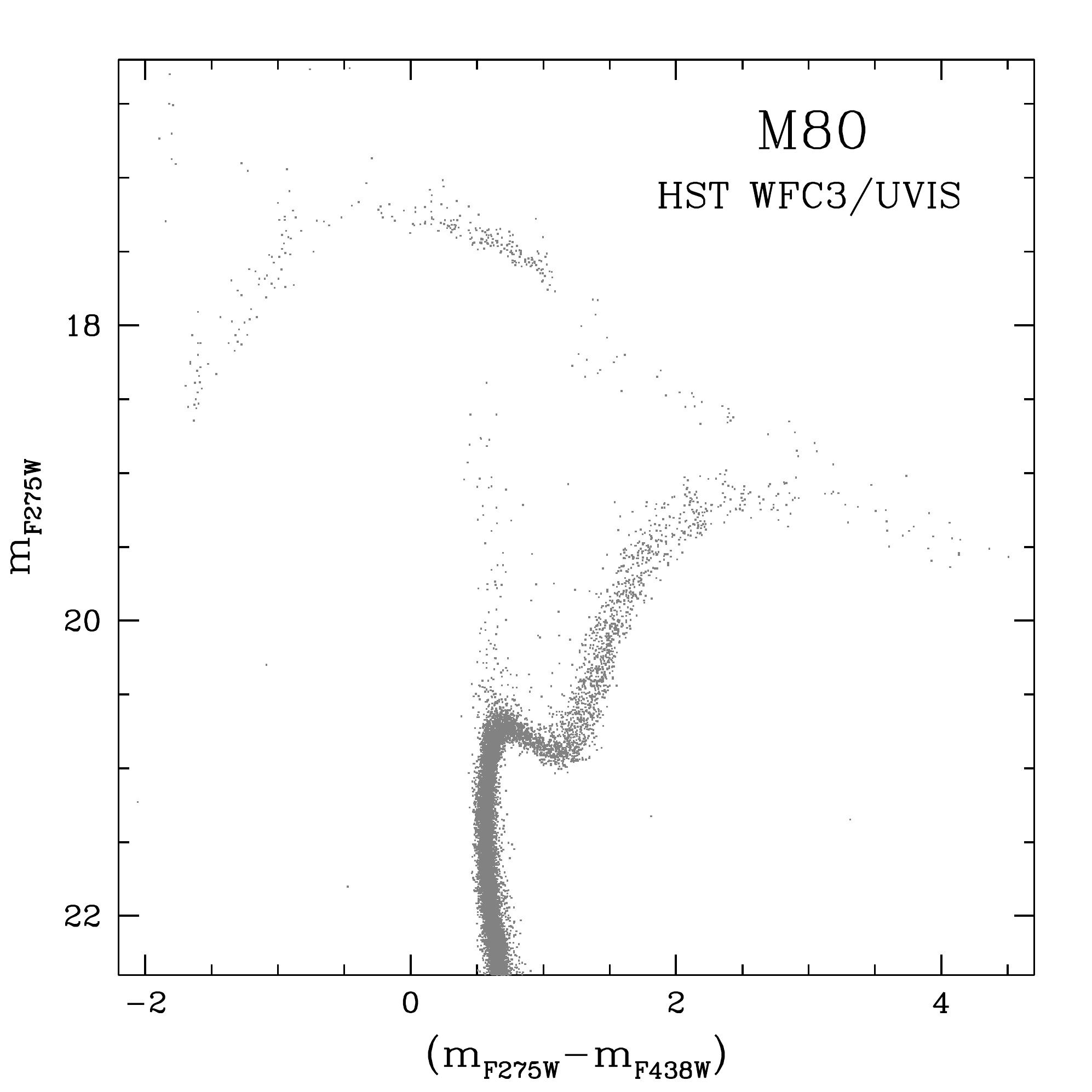}
\caption{($m_{F275W}, m_{F275W}-m_{F438W}$) CMD of the stars in the HRC-WFC3 field of view.}
\end{figure}

The photometric analysis has been performed
independently on each chip by using {\tt DAOPHOT IV} \citep{stetson87}.
Tens of bright and isolated stars have been selected in each frame to
model the point-spread-function. For the following analysis 
we took advantage of the reduced crowding
conditions at UV wavelengths by using the so-called ``UV-route'' approach, which is particularly effective for very dense systems like M~80
\citep[see for example][]{ferraro98,ferraro03,dalessandro08,dalessandro09,raso17}.
Briefly, a first star list has been obtained for each image by
independently fitting star-like sources above the $5\sigma$ level from
the local background. Starting from these preliminary catalogues, we then created 
two UV master-lists. For the most crowded 
innermost regions, we created a master-list by using stars detected in the HRC F250W 
image, which provides a better sampling than the WFC3 and helps to properly
resolve the central regions. In the WFC3 field of view (FOV) complementary to HRC, the master-list has been obtained 
by using stars detected in at least half of the F275W images. 
At the corresponding positions of stars in the combined WFC3+HRC master-lists, a fit was
forced in all the available images using {\tt DAOPHOT/ALLFRAME} \citep{stetson94}. For each star thus
recovered, multiple magnitude estimates obtained in each chip were
homogenized by using {\tt DAOMATCH} and {\tt DAOMASTER}, and their weighted mean and
standard deviation were finally adopted as star magnitude and photometric
error.

\begin{table}
\centering\begin{tabular}{ c  c  c  c }
\hline
\hline
Instrument & Filter & ${\rm t}_{{\rm exp}}$ (s) & Proposal ID/PI\\
\hline
\multirow{1}{*} {\small ACS/HRC}      & { \scriptsize F250W} &  { \scriptsize $1 \times 2348$}    & \multirow{1}{*} { {\small GO-10183/Knigge}}\\

\hline
\multirow{2}{*} {{ \small ACS/WFC}}        & { \scriptsize F606W}  & { \scriptsize $4 \times 603 + 1 \times 60$}  & \multirow{2}{*}{{ \small GO-10775/Sarajedini}}\\ 
\multirow{2}{*} {}      		   & { \scriptsize F814W}  & { \scriptsize $4 \times 60$}	      	& \multirow{2}{*}{}\\
\hline

\multirow{3}{*}  {\small WFC3/UVIS} &   { \scriptsize F275W} & { \scriptsize $10\times 855$}       &  \multirow{3}{*}{{ \small GO-12605/Piotto}}\\
\multirow{3}{*}  {}                 &   { \scriptsize F336W} & { \scriptsize  $5 \times 657$}      & \multirow{3}{*}{}\\
\multirow{3}{*}  {}                 &   { \scriptsize F438W} & { \scriptsize $5 \times 85$}        & \multirow{3}{*}{}\\

\hline
\multirow{3}{*}  {\small WFPC2}     &   { \scriptsize F160BW} & { \scriptsize $4\times 900$}       &  \multirow{3}{*}{{ \small GO-5903/Ferraro}}\\
\multirow{3}{*}  {}                 &   { \scriptsize F336W} & { \scriptsize  $4 \times 600$}      & \multirow{3}{*}{}\\
\multirow{3}{*}  {}                 &   { \scriptsize F555W} & { \scriptsize $4 \times 23 + 2 \times 2$}        & \multirow{3}{*}{}\\
\hline
\hline
\end{tabular}
\caption{Summary of the HST data-sets used in this work.}
\label{dataset}
\end{table}

Instrumental magnitudes were reported to the VEGAMAG photometric
system by using equations and zero-points reported in the dedicated HST web pages. 

By using the F606W and F814W magnitudes and following the approach described in \citet{milone12a}
we estimated the differential reddening within the WFC3/UVIS FOV. 
Briefly, we rotated the optical CMD onto a photometric reference frame where the abscissa is parallel to the reddening vector. 
In this reference frame we then defined a fiducial line along the brighter portion of the MS and 
calculated the distance ($d_{\rm redd}$) from it along the reddening direction of stars in that magnitude range, which were used as reference stars. 
For each star in our catalog we computed the average value of $d_{\rm redd}$ of the 30 closest reference stars, 
which was then used to derive the reddening variation to associate to that star. 
We find a maximum variation 
$\Delta(E(B-V))\sim 0.045$ mag within the entire FOV. The following analysis is based on differential-reddening corrected magnitudes. 

Instrumental coordinates were reported onto the absolute system 
($\alpha$, $\delta$) by using the stars in
common with the ACS Survey of Galactic Globular Clusters
catalog\footnote{The catalog is available on-line at the following link: \url{https://www.astro.ufl.edu/~ata/public_hstgc/}} as secondary astrometric
standards and the cross-correlation software CataXcorr.

The resulting ($m_{F275W}, m_{F275W}-m_{F438W}$) CMD for stars in the WFC3 FOV is shown in Figure~2.
The CMD shows clearly the presence of a well extended and
multi-modal Horizontal Branch (e. g. \citealt{ferraro98,dalessandro11b}) and of a large populations of BSSs \citep{ferraro03} describing 
an almost vertical sequence extending for
about 2 magnitudes in F275W. 

For the analysis of the horizontal branch, we used the HST WFPC2 catalog (see details about the data-set in Table~1)
published by \citet{ferraro98}.
For the present analysis magnitudes were reported to the VEGAMAG photometric system and corrected for {\tt CTE} by means of the prescription by \citet{dolphin00} and updated equations listed in the Dolphin's 
website\footnote{\url{http://purcell.as.arizona.edu/}}. We remind that the adoption of the F160BW 
far-UV band and the use of the ($m_{F160BW}, m_{F160BW}-m_{F555W}$) CMD is key for clusters with an extended horizontal branch such as M~80. In fact, 
in this diagram, the hottest HB stars are the most luminous and lie along almost horizontal sequences whose luminosity is very sensitive to the initial Y abundance irrespective of the precise value of the stellar mass.

\section{Multiple populations along the Red Giant Branch}

Figure~3 (left panel) shows the ($m_{F336W}, C_{275,336,438}$) pseudo-color diagram of M~80, where  
$C_{275,336,438}=((m_{F275W}-m_{F336W})-(m_{F336W}-m_{F48W})$).
This UV color combination has been used to characterize the LE-MP properties in a large sample of GCs within {\it The HST UV Legacy Survey of globular clusters} \citep{piotto15,milone17}. The combination efficiently highlights the presence of LE-MPs since 
it traces simultaneously the strength of the OH, NH and CH molecular bands. 
We verticalized the distribution of RGB stars in the magnitude range $17.7<m_{F336}<19.4$ in the 
($m_{F336W}, C_{275,336,438}$) diagram with respect to 
to two fiducial lines at the bluest and reddest color edges of the RGB (left panel of Figure~3), 
which have been drawn to include the bulk of RGB stars in the considered magnitude interval.
The derived verticalized color distribution ($\Delta_{C_{275,336,438}}$) appears to shows three main components (middle panel of Figure~3)
that we fit with Gaussian Mixture 
Models (GMMs). The result of the fit is shown in the middle panel of Figure~3. Three components can be identified that we classified as FG (red), SG$_{\rm INT}$ (intermediate SG - green) and SG$_{EXT}$ (extreme SG - blue) for increasing values of $\Delta_{C_{275,336,438}}$ as the RGB is populated by stars increasingly enhanced in N moving from red to blue colors. From the areas under the gaussian functions, 
we computed the number ratios among different sub-populations.
In particular, we find that
$N_{\rm FG}/N_{\rm TOT}=0.395\pm0.023$ (where $N_{\rm FG}$ is the number of FG stars and $N_{\rm TOT}$
the total number of stars selected along the RGB), which is compatible within the errors with that obtained by 
\citet{milone17} ($N_{\rm FG}/N_{\rm TOT}=0.351\pm0.029$).

Because of the combination of intrinsic color spreads and photometric errors, some degree of overlap is present among the selected sub-populations. To minimize 
possible contamination among them, in the analysis described in the next Sections we will use only stars with a probability $P>85\%$ to belong to a given population.

\begin{figure*}
\plotone{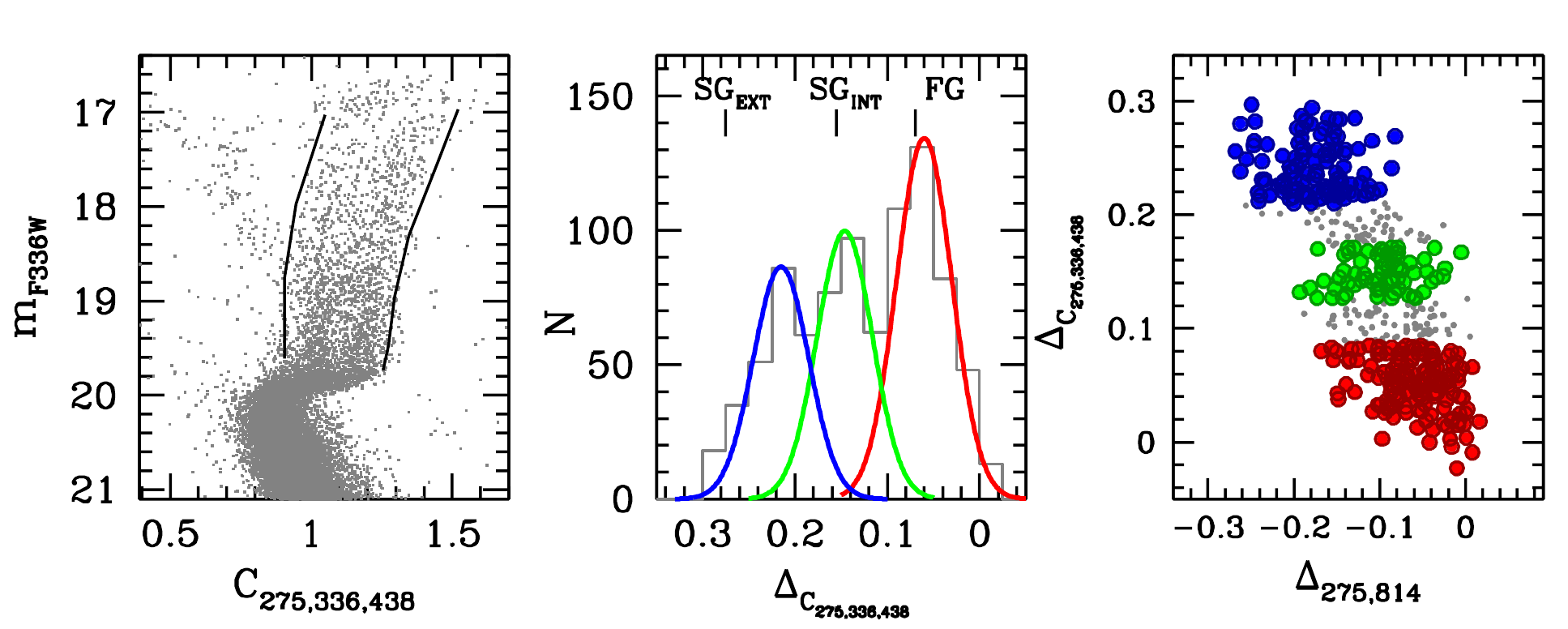}
\caption{Left panel: ($m_{F336W}, C_{275,336,438}$) pseudo-color diagram of M~80. 
The middle panel shows the $\Delta_{C_{275,336,438}}$ distribution of RGB stars in the magnitude range $17.7<m_{F336}<19.4$. 
Three main components can be identified (see Section~3) that we defined as 
FG, SG$_{INT}$ and SG$_{EXT}$ in red, green and blue respectively. Right panel: ($\Delta_{275,814}$, $\Delta_{C_{275,336,438}}$) color-color diagrams of the selected LE-MPs. Colored points represent stars with $P>85\%$ to belong to one of the three populations.}
\end{figure*} 

The right panel of Figure~3 shows the so-called ``chromosome map'' ($\Delta_{275,814}$, $\Delta_{C_{275,336,438}}$).
$\Delta_{275,814}$ was obtained by verticalizing  
the distribution of RGB stars in the ($m_{F814W}, m_{F275W}-m_{F814W}$) CMD 
with respect to the fiducial lines at the blue and red edges of the RGB. 
The three sub-samples our analysis is focused on are clearly separated in this diagram.

\begin{figure}
\plotone{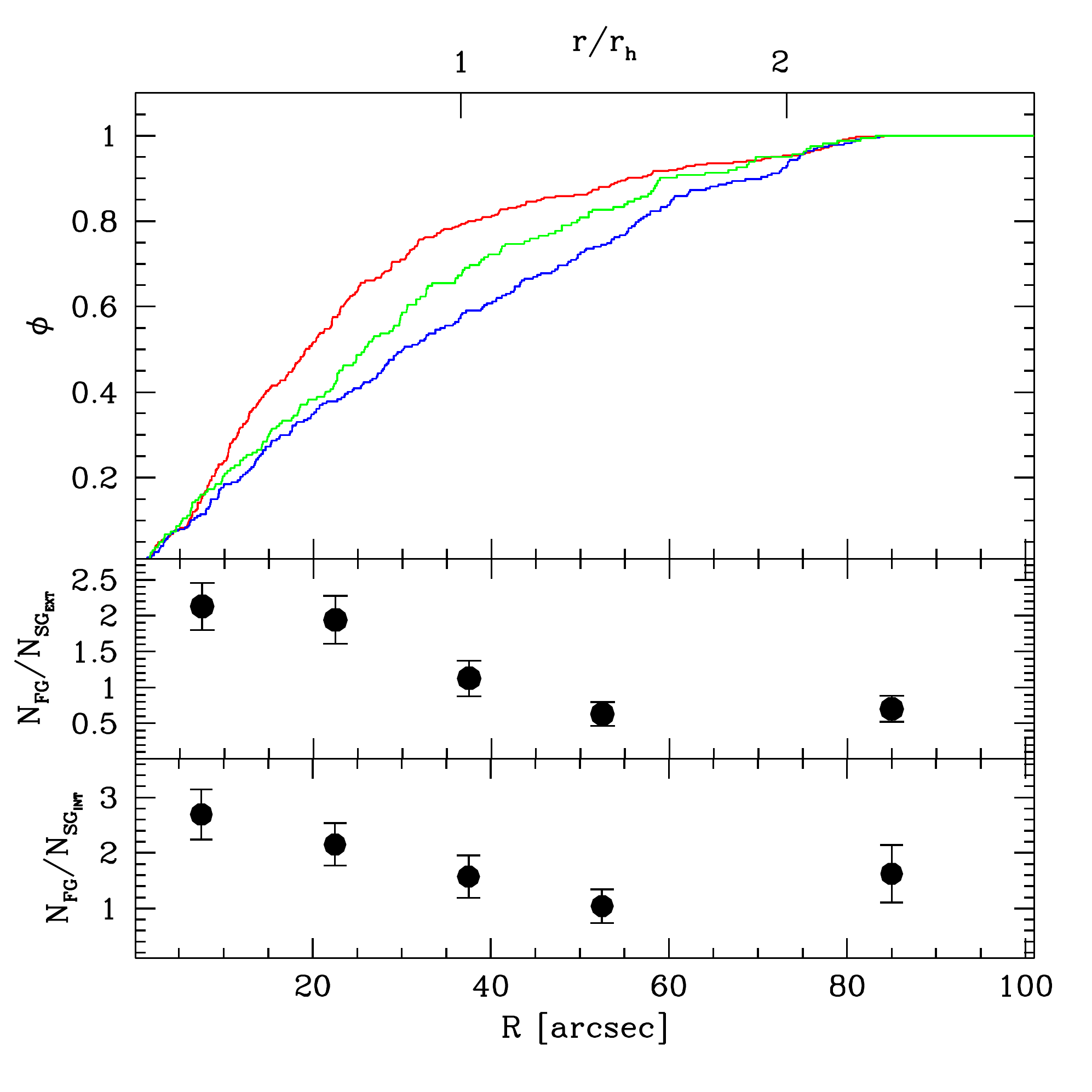}
\caption{Upper panel: cumulative radial distribution of FG (red), SG$_{\rm INT}$ (green) and SG$_{\rm EXT}$ (blue) stars selected as detailed in Section~3. Lower panels: N$_{FG}$/N$_{SG_{\rm INT}}$ and N$_{FG}$/N$_{SG_{\rm EXT}}$ as a function of the distance from the cluster center.}
\end{figure}

\section{Radial distribution of multiple populations}  

To study the radial distribution of the LE-MPs of M80, we first derived the center of the cluster 
($C_{\rm grav}$). 
$C_{\rm grav}$ was determined by using an iterative procedure that averages the positions $\alpha$ and $\delta$ of stars in a defined magnitude range and lying within a given 
distance from a first guess center. At each iteration star distances are re-calculated respect to the center obtained in the previous iteration until convergence is reached \citep[see for example][for more details]{dalessandro13,cadelano17}.
We used as starting guess center the one found by \citet{goldsbury10}.
To avoid spurious and incompleteness effects, we performed various measures of $C_{\rm grav}$ using stars in different magnitude intervals with lower limits in the range $20.9<m_{F275W}<21.3$ and distance ($d$) from the 
cluster guess center in the range $0\arcsec<d<30-40\arcsec$.
The resulting $C_{\rm grav}$ is the average of all these measures, and it is located at ($\alpha=16^{h}17^{m}2.481^{s} $, $\delta=-22^{\circ}58\arcmin 34.098\arcsec$), with an uncertainty of about $0.17\arcsec$. The newly determined center is located at $\sim1\arcsec$ from that obtained by \citet{goldsbury10}.

We analyzed the radial distributions of FG, 
SG$_{\rm INT}$ and SG$_{\rm EXT}$ stars with respect to $C_{\rm grav}$ within the entire WFC3 FOV. 
The WFC3 FOV extends out to $\sim 2.5\times r_h$ (where $r_h$ is the half-light radius -
$r_h=36.6\arcsec$; \citealt[][2010 edition]{harris96}) and samples $\sim 80\%$ of the total light of the cluster.

Figure~4 shows that the cumulative radial distributions of the three sub-populations 
are significantly different.
Surprisingly we find that FG stars are more centrally concentrated than 
the other two sub-groups with the largest difference with SG$_{\rm EXT}$.  
According to the K-S test,
the probability that the FG and SG$_{\rm EXT}$ are extracted from the same parent population is $P\sim10^{-7}$, while $P\sim3\times10^{-3}$ for FG and SG$_{\rm INT}$ stars.
This result is qualitatively similar to what was found in M~15 by \citet[][but see \citealt{nardiello18} for a recent analysis 
of M~15 in which the trend found by Larsen et al. is not confirmed]{larsen15} and it is unexpected
in the context of theories of formation and dynamical evolution of LE-MPs available so far.

In the lower panels of Figure~4 the variations of the $N_{FG}/N_{SG_{EXT}}$ and $N_{FG}/N_{SG_{INT}}$
ratios as a function of the distance from $C_{\rm grav}$ are also shown. 
FG stars are $\sim 2$ and $\sim1.5$ times more numerous than
SG$_{\rm EXT}$ and SG$_{\rm INT}$ stars respectively in the innermost regions ($\sim30\arcsec$), then the $N_{FG}/N_{SG_{EXT}}$ ratio monotonically decreases 
reaching a minimum value of $\sim 0.6$, while the $N_{FG}/N_{SG_{INT}}$ ratio seems to have (with moderate significance) a bimodal distribution 
which reaches a minimum value
($N_{FG}/N_{SG_{INT}}\sim0.8$) at $d\sim50\arcsec$ and then rises again outward.

\begin{figure}
\plotone{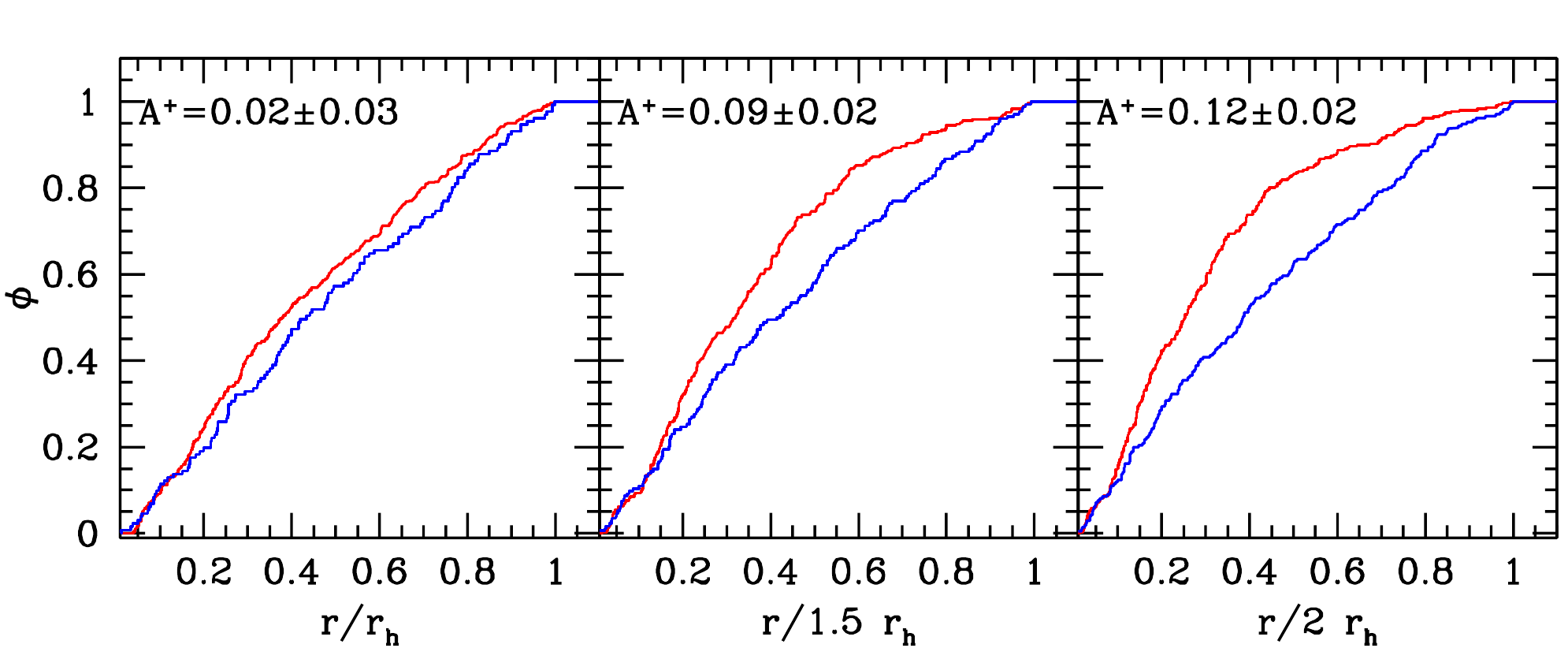}
\caption{Cumulative radial distributions of FG (red) and SG$_{\rm EXT}$ (blue) stars as obtained for 1.0, 1.5 and 2.0 $r_h$ (moving from the left to the right panel) from $C_{grav}$. Derived values of $A^+$ are reported in each panel.
}
\end{figure} 

In order to verify the effect of photometric incompleteness on the radial distributions shown in Figure~4,
we performed a large number of artificial star experiments following the approach described in \citet[][see also \citealt{dalessandro15}]{dalessandro11a}.
We find that the three sub-samples have photometric completeness $C>95\%$ in the entire area 
sampled by our analysis, therefore their 
radial distributions are virtually unaffected by completeness radial variation effects.

It has been shown that the difference in segregation between different sub-populations can be quantitatively estimated 
by using the parameter $A^+$. 
It was first introduced by \citet[][see also \citealt{lanzoni16}]{alessandrini16} as a mass-segregation indicator for BSS and it is simply defined as the area enclosed between the cumulative radial distributions of two samples of stars. $A^+$ has the advantage of not requiring binning of data and allowing an easy and direct comparison among different clusters.

Here we have used the cumulative radial distribution of SG$_{\rm EXT}$ stars as reference and we adopted the following definition of $A^+$:

\begin{equation}
A^+(x)=\int_{x_{min}}^x \phi_{FG}(x^{'})-\phi_{SG_{EXT}}(x^{'})dx^{'}
\end{equation}

where $\phi_{\rm POP}(x)$ is the cumulative radial distribution, $x=(d/d_{max}$) and $d_{max}$ is the maximum distance (expressed in units of $r_h$) within which $A^+$ is calculated.
With such a definition star distances are always comprised in the range 0-1 and
a more segregated FG sub-population leads to positive A+ values.  Taking advantage of the wide area coverage, 
we have explored the radial variation of A+ by measuring it for three distances: 1.0, 1.5 and 2.0 $r_h$ from $C_{grav}$ (Figure~5). 
With these prescriptions, we obtain $A^+(1 r_h)=0.02\pm0.02$, $A^+(1.5 r_h)=0.08\pm0.02$ and $A^+(2.0 r_h)=0.10\pm0.01$ when
FG and SG$_{EXT}$ sub-populations are considered and 
$A^+(1 r_h)=0.0\pm0.02$, $A^+(1.5 r_h)=0.03\pm0.02$ and $A^+(2.0 r_h)=0.04\pm0.02$ for FG and SG$_{INT}$ stars. 
Uncertainties in $A^+$ have been estimated by applying a jackknife bootstrapping technique \citep{lupton93}. 
$A+$ is recomputed by leaving out one different star from the two considered samples each time. In this way, given a sample of N stars,
 we end up with N estimates of $A^+$ computed on samples of N-1 stars. The uncertainty on $A^+$ is therefore 
$\sigma_{A^+}=\sqrt{N-1} \times \sigma_{sub}$, where $\sigma_{sub}$ is the standard deviation of the $A^+$ distribution derived from N subsamples.

\section{N-body simulations}

To understand to what extent the observational results presented in the previous sections represent an anomaly 
in the context of the current theoretical models for the formation and evolution of LE-MPs, 
we have explored the evolution of the spatial distribution of FG and SG stars in two sets of N-body 
simulations following the long-term evolution of the structural properties of multiple-population clusters.

Our simulations start with 50000 stars equally split between FG and SG stars and have been run using the GPU-accelerated version of the code NBODY6 \citep{aarseth2003,nitadori2012}.
Both populations follow the spatial distribution of a \citet{king66} model with central dimensionless potential equal to $W_0=7$, but different values of $r_h$. 
In particular, we have explored the evolution of two systems: one in which the SG population's $r_h$ ($r_h^{SG}$) is about 5 
times smaller than $r_h^{FG}$ and one in which $r_h^{SG}$ is about ten times smaller than $r_h^{FG}$.  
The stellar masses are assigned assuming a \citet{kroupa01} initial mass function evolved to an age of 11.5 Gyr
(the software McLuster -- \citealt{kuepper11} -- has been used to setup the initial conditions).
The cluster is tidally limited and assumed to move on a circular orbit in the external potential of a host galaxy modeled as a point mass.
As cluster evolves, it loses mass due to the combined effect of two-body relaxation and tidal truncation.

We emphasize that the simulations presented here are still simplified and not tailored to provide specific detailed models for M80.
Our goal is to illustrate the general evolution of the spatial distributions of FG and SG stars using the $A^+$ parameter 
and provide a quantitative measure of the dynamical significance of the effect found in our observational analysis.

In panels a) and b) of Figure 6 we show the time evolution of the A+ parameter calculated for FG and SG stars with masses $M_{FG}=M_{SG}$ between $0.80M_{\odot}$ and $0.85M_{\odot}$ (as appropriate for RGB stars in an old GC). 
Time is expressed in units of the instantaneous half-mass relaxation time, $t_rh$. The A+ parameter is determined within three projected distances from the center: 1, 1.5, and 2 $r_h$ (dark turquoise, pink and blue curves respectively), with $r_h$ being the half-mass radius.

The systems start with the SG concentrated in the cluster's innermost regions, which, 
considering the definition of $A^+$ in Section~4, corresponds to negative values of $A^+$. 
Panel a) shows the results for simulations starting with $r_h^{FG}=5\times r_h^{SG}$, while panel b) those for 
$r_h^{FG}=10\times r_h^{SG}$.
As the systems evolve and the two populations gradually mix, $A^+$ increases and evolves towards the value of $A^+$  corresponding to a mixed cluster ($A^+=0$). 

It is interesting to note that
in our simulations $A^+$ can reach positive values 
for dynamically old and mixed systems (at, approximately, $t/t_{rh}>5-10$). 
However, we stress that they should not be interpreted as real inversions of the radial distribution, but only as small fluctuations not corresponding to any secular evolution.

We compare our N-body simulations with the observed values of $A^+$ (grey symbols in Figure~6) by
focusing now on the comparison between FG and SG$_{\rm EXT}$, as they show the most discrepant and significant behavior.
For simplicity in the following we will refer to them as FG and SG respectively.
Clearly observations appear to deviate significantly from the range of values spanned by $A^+$ in our simulations when equal mass FG and SG are assumed and 
they are not compatible with the expected fluctuations.  
This suggests that the interpretation of the radial distribution of LE-MPs in M80 requires some additional dynamical ingredient.

\begin{figure}
\plotone{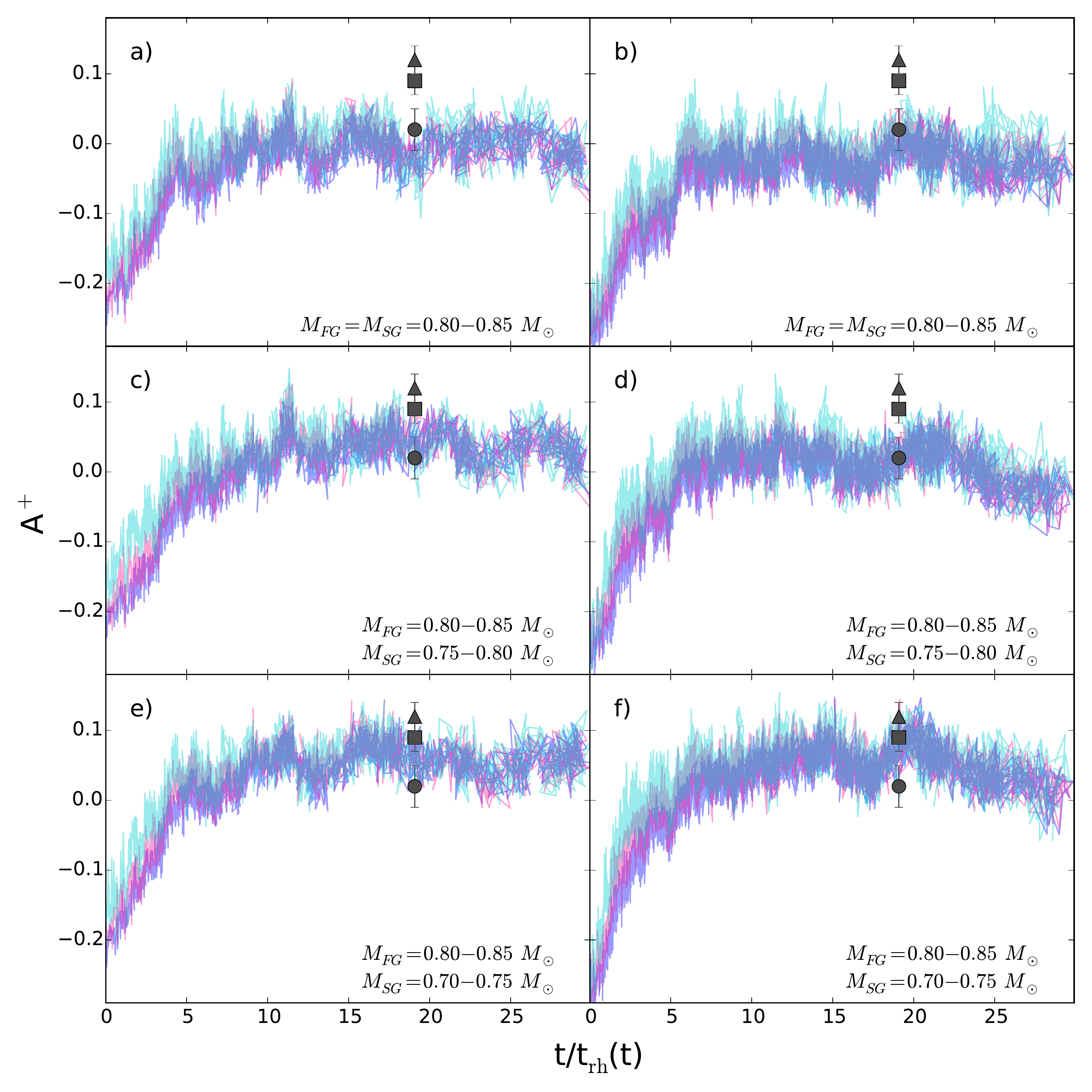}
\caption{Time evolution of the $A^+$ parameter for the simulations starting with $r_h^{FG}=5\times r_h^{SG}$, (left hand panels) and those for $r_h^{FG}=10\times r_h^{SG}$ (right hand panels). 
The two panels in the top row (a) and b)) show the evolution of $A^+$ calculated using FG and SG stars with masses 
between 0.80 and 0.85 $M_{\odot}$. The panels in the middle row (c) and d)) show the evolution of $A^+$ calculated using FG stars with masses between 0.80 and 0.85 $M_{\odot}$ and SG stars with masses between 0.75 and 0.80 $M_{\odot}$. The panels in the bottom row (e) and f)) show the evolution of $A^+$ calculated using FG stars with masses between 0.80 and 0.85 $M_{\odot}$ and SG stars with masses between 0.70 and 0.75 $M_{\odot}$.
In each panel the three lines show the time evolution of $A^+$ calculated within a distance from the cluster centre equal to 1 (dark turquoise), 1.5 (pink) and 2 (blue) projected half-mass radii.
Time is normalized to the instantaneous half-mass relaxation time. The three grey dots show the observational values calculated within 1, 1.5 and 2 (circle, square and triangle respectively) half-mass radii for the FG and SG sub-populations in M80 (see Section 4).}
\end{figure} 

First we note that even if we assumed 
that FG stars formed more concentrated than SG stars
the two sub-populations would be mixed for a cluster with the dynamical age of M~80 ($t/t_{r_h}\sim20$).
No matter what the details of the formation history of the cluster are and independently of which population was initially more concentrated,
a primordial origin for the observed relative spatial distributions of the FG and SG populations observed in M80 seems difficult 
to reconcile with the cluster's long-term dynamical evolution.

We now investigate the possibility that the different radial distributions are due to
a different average stellar mass of the two sub-populations. 
This would indeed be the case if the two populations are characterized by different helium abundances. 
This option was considered also by \citet{larsen15} to explain the inner inversion in the radial profile 
of the SG-to-FG number ratio found in M15. 
However, in that case the mass difference required was too large 
and inconsistent with the observations.

In Figure~6 panels from c) to f) we show the time evolution of $A^+$ calculated for FG stars with masses in the range $0.80-0.85 M_{\odot}$ 
and SG stars with two different stellar mass ranges: $0.75\leq m/M_{\odot} \leq 0.80$ (panels c) and d)) and $0.70\leq m/M_{\odot} \leq 0.75$ (panels e) and f)).
In these two cases the difference between the mean mass of FG stars and that of SG stars is, respectively, 
equal to about $0.05 M_{\odot}$ and  $0.1 M_{\odot}$. These mass differences correspond to an
enhancement in the helium abundance of SG stars of $\Delta Y \sim 0.05-0.06$ (see Section~6).

Figure~6 (panels from c) to f)) clearly demonstrates that when SG stars are assumed 
to be slightly less massive than FG stars, 
$A^+$ evolves towards positive values corresponding to a configuration in which the less massive SG stars are less concentrated than the more massive FG stars. 
Even the small mass difference we considered in our simulations is sufficient to significantly shift the values 
reached by $A^+$ late in the cluster evolution towards positive values. 
Such an effect brings the simulations in good agreement with the observational values of $A^+$ for M80.
As a consequence, a different average mass between FG and SG$_{EXT}$ stars appears to be a viable solution to reproduce the unexpected radial distribution in this system.

Another possible explanation was proposed by \citet{henault15} who suggested that the preferential 
disruption of SG binaries (see e.g. \citealt{vesperini11,hong15,hong16}) 
might lead to the higher central 
concentration of FG stars relative to SG stars observed in the central regions.
No models or simulations were carried out in \citet{henault15} to test this hypothesis and it is unclear whether this is actually a viable option.

The solution based on mass differences seems to be dynamically robust.
However, we note that this possibility should be further tested by additional studies of the spatial distributions in clusters 
with extreme populations characterized by non-negligible differences in helium. 
As shown by our simulations however, also in these cases a cluster needs to be dynamically 
old before the dynamical consequences of differences in mass manifest themselves in the spatial distribution of the two populations. 
Moreover the extent of this effect may also depend on the radial coverage of the observational data used to calculate the $A^+$ parameter. 

\begin{figure}
\plotone{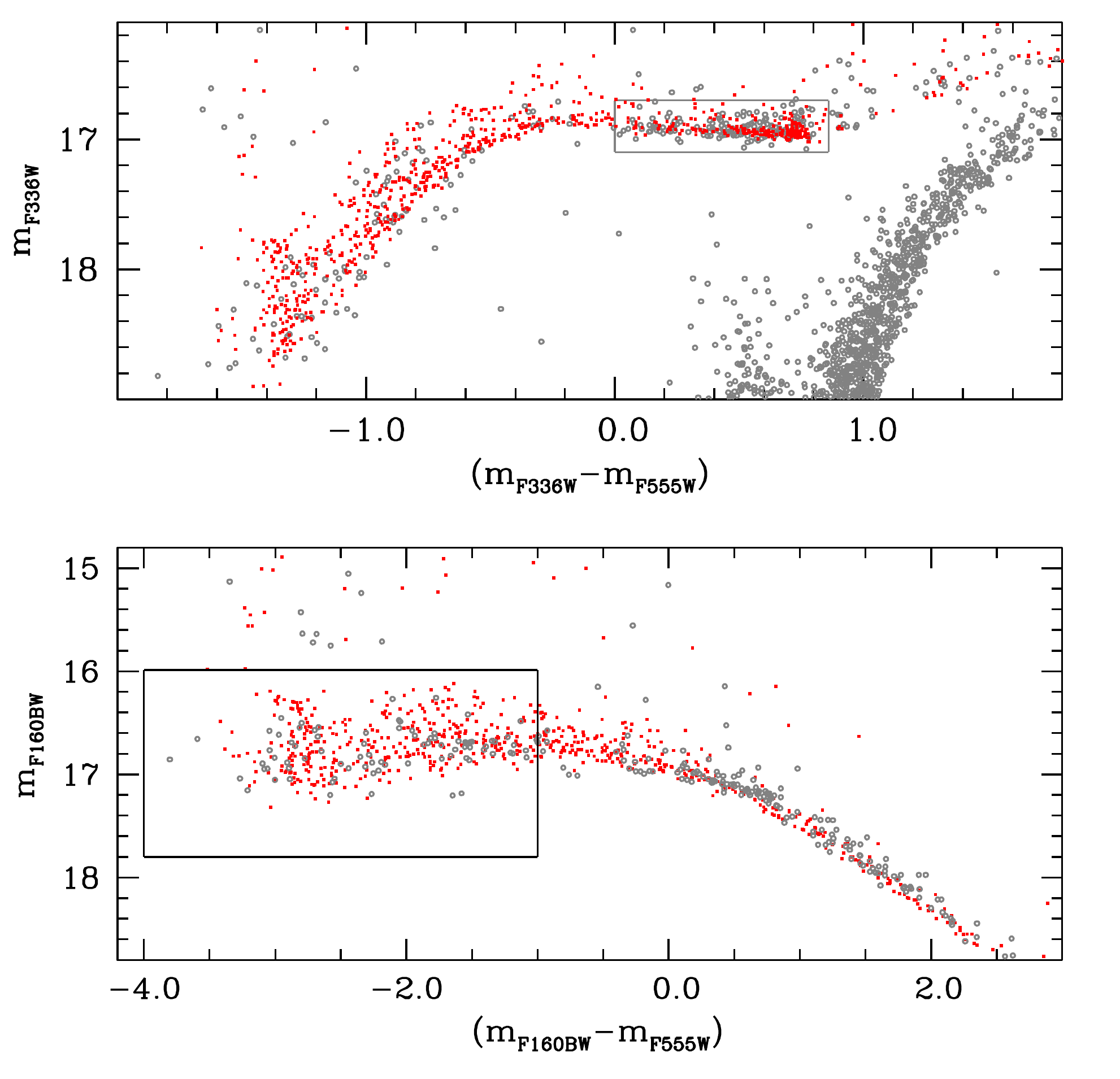}
\caption{Observed ($m_{F336W}, m_{F336W}-m_{F555W}$) and ($m_{F160BW}, m_{F160BW}-m_{F555W}$) CMDs (upper and bottom panels respectively) 
of the HB stars of M~80 (black open circles). Black boxes define the red part of the HB (upper panel) and the blue one (bottom panel) 
where the HB analysis has been performed. Best-fitting synthetic HB (red points) is overplotted for comparison. }
\end{figure}

\section{Observational constraints on He abundance variations}

In this Section we investigate whether the average He (mass) difference between LE-MPs inferred from the dynamical study (Section~5) is compatible with 
He-abundance-sensitive photometric features in the CMD. To this aim we use a proper comparison between theoretical models and the 
observed morphology of the horizontal branch (HB).

To constrain the He spread necessary to match the observed HB morphology of M~80, we employed the same approach described in detail by  
\citet{dalessandro11b,dalessandro13}, which is based on a comparison between synthetic HB calculations and UV-optical CMDs. 
For the portion of HB redder (cooler) than the RR Lyrae instability strip, this comparison is performed with stars within the black box in the ($m_{F336W}, m_{F336W}-m_{F555W}$) CMD (Figure~7, upper panel), while for bluer stars we take advantage of the reduced bolometric corrections  at UV bands by using stars located within the box in the ($m_{F160BW}, m_{F160BW}-m_{F555W}$) diagram (Figure~7, lower panel). In both boxes stars are distributed almost horizontally in the corresponding CMDs.
As a consequence in these regions the effect of He variations is more pronounced.

\begin{figure}
\plotone{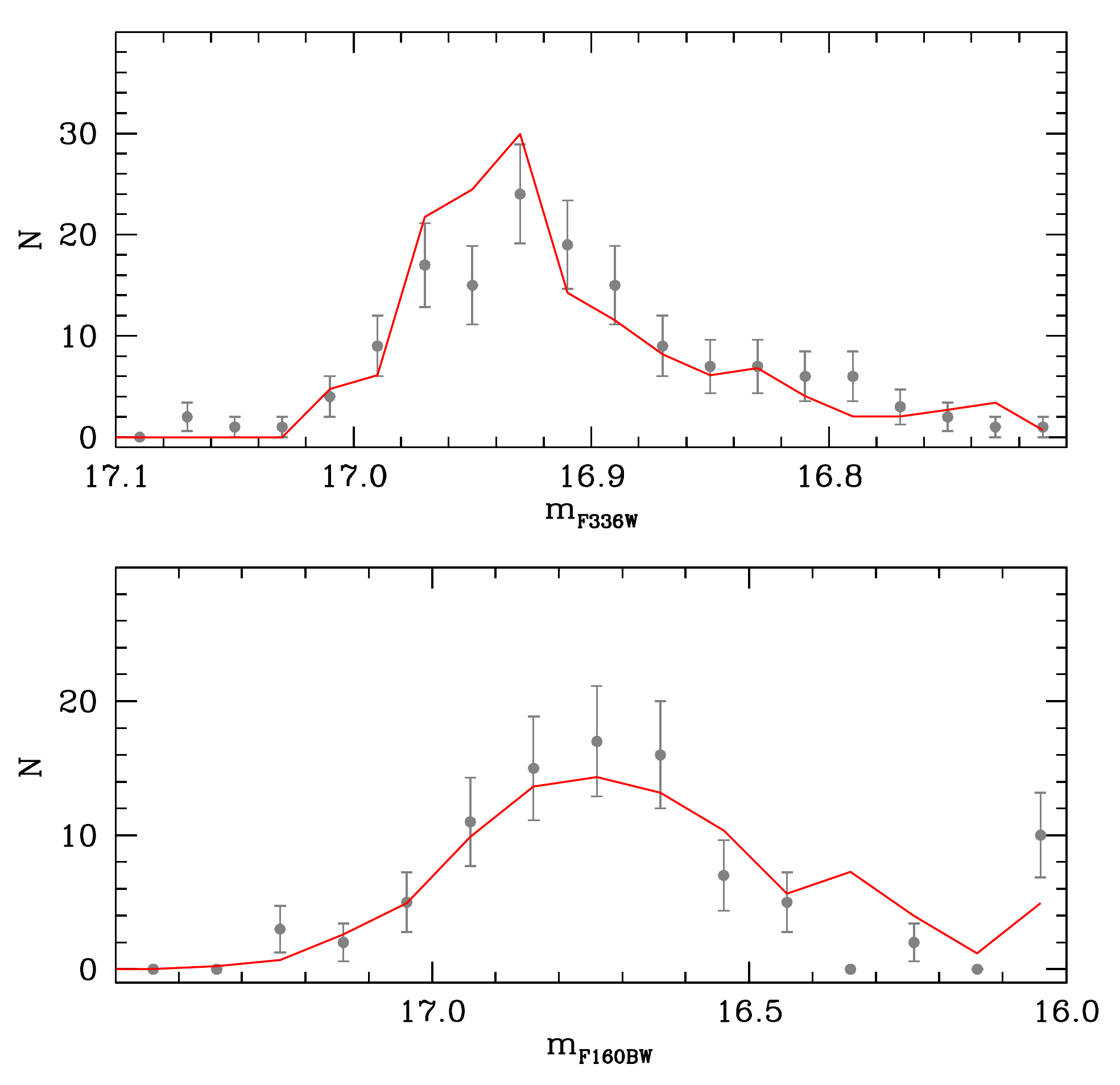}
\caption{Comparison between observed (grey points) and theoretical best-fit (red line) star counts for the HB population, as a function of $m_{F336W}$ and $m_{F160BW}$ magnitudes.}
\end{figure}  

Our HB simulations require the specification of four parameters: the minimum value of $Y$ ($Y_{\rm min}$, fixed to $Y$=0.246), 
the range of He abundances ($\Delta Y$), the 
mean value of the mass lost along the RGB ($\Delta M$) and the spread around it ($\sigma(\Delta M)$).
Our synthetic HB code first draws randomly a value of $Y$ (we assume for simplicity a uniform probability distribution) 
and determines the initial mass of a star at the RGB tip ($M_{TRGB}$) from interpolation among a 
set of BaSTI $\alpha$-enhanced HB tracks and isochrones \citep{pietrinferni06} with [Fe/H]=$-$1.62, and an age of 12~Gyr.
The mass of the corresponding object evolving along the HB ($M_{HB}$) is then given by $M_{HB}=M_{TRGB}-\Delta M$ plus a random Gaussian perturbation $\sigma(\Delta M)$.
The WFPC2 magnitudes of the synthetic star are then determined according to its position along the HB track with appropriate mass and Y, obtained by 
interpolation among the available set of HB tracks, after an evolutionary time $t$ has been randomly extracted, 
assuming that stars reach the zero age horizontal branch (ZAHB) at a constant rate. We employed therefore a flat probability
distribution for $t$ ranging from zero to ${\rm t_{HB}}$, where  ${\rm t_{HB}}$ denotes the time
spent from the ZAHB to the He-burning shell ignition along the
early asymptotic giant branch. The value of ${\rm t_{HB}}$ is set by the mass with the longest
lifetime (the lowest masses for a given $Y$ and $Z$). This implies that
for some synthetic object the randomly selected value of $t$ will be
longer than its ${\rm t_{HB}}$ or, in other words, that they have already evolved
to the next evolutionary stages. 
The derived synthetic magnitudes are finally perturbed with a Gaussian $1\sigma$ error determined from the data quality and reduction procedures. As done 
by \citet{dalessandro11b,dalessandro13}, we mimicked the effect of radiative levitation by applying bolometric corrections appropriate for [Fe/H]=0.0 
and scaled-solar mixture when the effective temperature is larger than 12000 K. Bolometric corrections and extinction effects have been calculated as in \citet{dalessandro11b}. 

\begin{figure}
\plotone{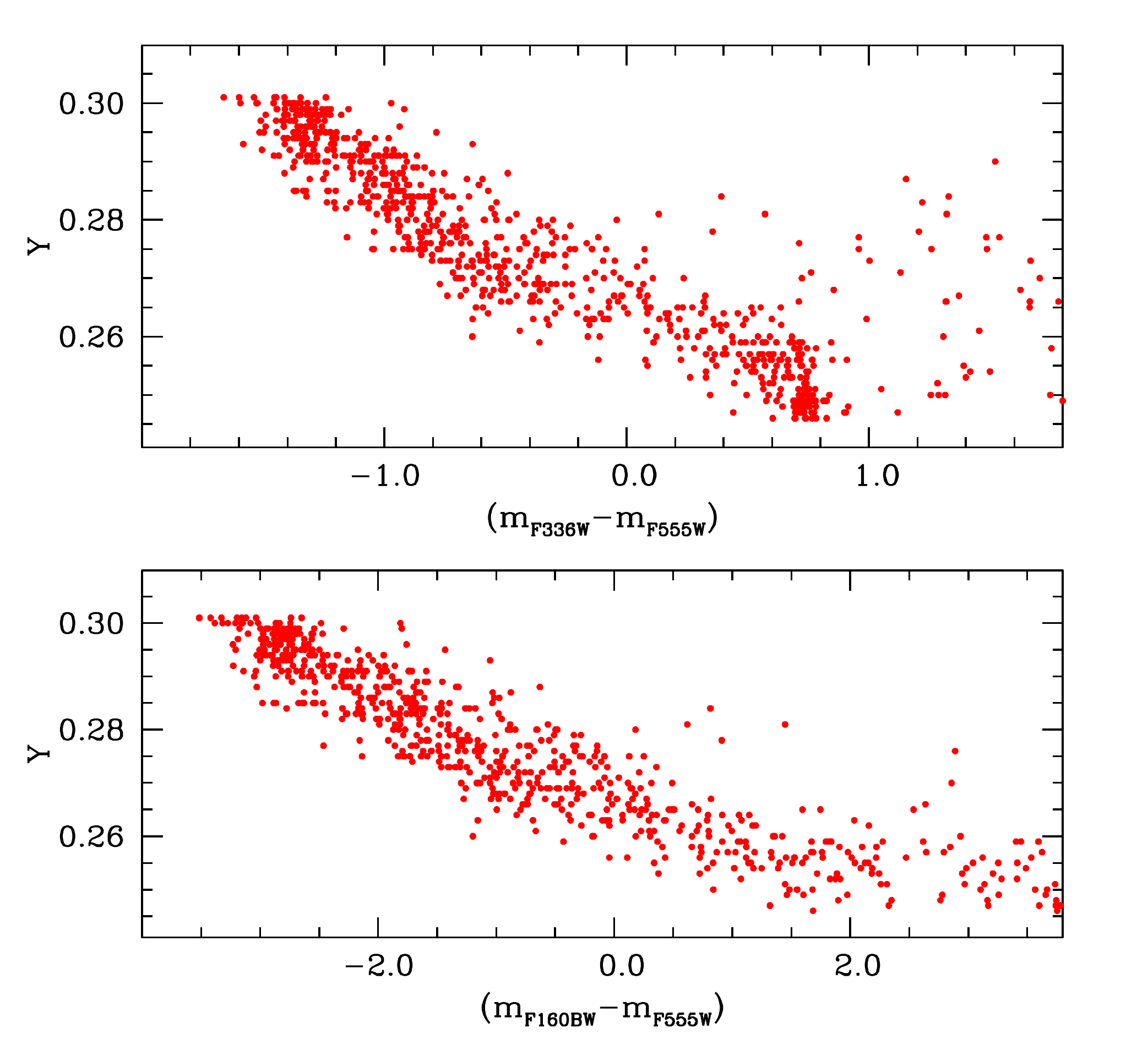}
\caption{$Y$ distribution as a function of the ($m_{F336W}-m_{F555W}$) and  ($m_{F160BW}-m_{F555W}$) colors along the HB of M80 for the best-fitting model.}
\end{figure} 

We adopted an average reddening of $E(B-V)=0.18$ \citep[see the 2010 version of the Harris globular cluster catalogue -- ][]{harris96}.
We then performed hundreds of HB simulations\footnote{The number of stars in each simulation is much higher than the observed sample, to reduce the effect of statistical number fluctuations on the
synthetic HB} by varying the distance modulus and exploring adequate ranges of $\Delta Y$, $\Delta M$ and $\sigma(\Delta M)$ to reproduce 
both the lower envelope --defined as the faintest magnitude bin
where the star counts are at least 2$\sigma$ above zero-- of the observed magnitude distributions in both the adopted boxes (Figure~7), and the mean 
values (and 1$\sigma$ dispersion) of the magnitudes in both boxes, with an accuracy better than 0.01~mag.

We find that the observed HB is best reproduced by a true distance modulus $(m-M)_0$ = 15.14$\pm$0.05  (which is in good agreement with the literature (see e.g \citealt{harris96}))
a maximum He variation $\Delta Y$=0.055, a mass loss law given by $\Delta M$=0.170 + $2 \times $ ($Y$-0.246) and a Gaussian spread $\sigma(\Delta M)$=0.015.\footnote{The required total RGB mass loss $\Delta M$ increases with increasing $Y$. 
A $\Delta M$ constant with varying $Y$ would not be able 
to cover the full observed color ranges, for the range of $Y$ that matches the observed magnitude distributions.}  
The resulting synthetic CMDs are shown in Figure~7, 
Figure~8 shows a comparison between the best-fit model and the observed star count distributions for the HB population and
Figure~9 shows the derived distribution of $Y$ as a function of the ($m_{F336W}-m_{F555W}$) and ($m_{F160BW}-m_{F555W}$) colors. 

This analysis demonstrates that a range of He abundance reaching $\Delta Y=0.05-0.06$ is needed to reproduce the HB morphology of M~80 and its magnitude is in very good agreement with that obtained by using dynamical constraints (see Section~5).
We warn the reader that the exact values of the derived He abundances are subject to uncertainties mainly 
because of the assumptions made to simulate the effect of radiative levitation. 
However, we have verified that even assuming bolometric corrections appropriate for [Fe/H]$=+0.5$ to mimick radiative levitation, 
the quoted results change by $\Delta Y <0.01$.
With this caveat in mind, it is interesting to note that Figure~9 shows that a significant fraction of HB stars has $Y>0.28-0.29$ consistent with the number of SG stars observed along the RGB.
This observational result provides independent support to the interpretation of the LE-MP relative radial distribution provided in Section~5 and based on the average mass difference between FG and SG$_{EXT}$ stars.

\section{Summary and Conclusions}

We have combined HST optical and near-UV images to study the radial distribution of the LE-MPs in the massive globular cluster M~80 out to $\sim 2.5\times r_h$. 
At least three sub-populations differing in terms of light-element abundances can be clearly identified along the cluster's RGB. 
Surprisingly we find that FG stars are significantly more centrally concentrated than both SG$_{INT}$ and SG$_{EXT}$ sub-populations for the entire radial extension covered by our data, with the difference being more significant for FG and SG$_{EXT}$ stars.

Our puzzling findings seem to be in tension with models of formation of LE-MPs in GCs, which predict that SG stars form more centrally concentrated than FG ones and dynamical evolution gradually erase their initial differences eventually leading, for the most evolved clusters, to a complete spatial mixing. 
Indeed this result raises a number of questions concerning the possible role of cluster formation and dynamical history in determining the observed spatial distributions of FG and SG stars.

We used the $A^+$ parameter \citep{alessandrini16,lanzoni16} to provide a quantitative measure of the difference in the FG and SG spatial distributions observed in M80. 
This parameter, initially introduced to study the degree of segregation of BSS in globular clusters, is extremely useful in performing homogeneous and quantitative comparisons of the distributions of different types of stars with other clusters and dynamical models.

In order to shed light on the extent to which this observational result is actually an anomaly, we present the results of a set of N-body simulations following the time evolution of $A^+$ in multiple-population clusters.
We find that our simulations are not able to reproduce the observed values of $A^+$
when assuming that FG and SG stars have the same average mass.
In fact, in dynamically evolved systems like M~80, simulations predict FG and SG to be totally mixed irrespective of their original configuration.

We have explored the possibility that the different spatial distributions are caused by different average stellar masses of the FG and SG stars.
With this assumption, our N-body simulations are able to recover the range of observed values of $A^+$. In particular, we find that by assuming a small difference in the average mass of $\Delta M\sim 0.05-0.10 M_{\odot}$ (Figure 5 panels from c) to f)), the observed values of $A^+$ are nicely reproduced. We argue that the imposed mass of the two sub-populations can result from the different evolutionary time-scales of stars with a different He abundance of $\Delta Y\sim0.05-0.06$.
Interestingly, based on a detailed comparison between far-UV photometry and theoretical models, we find that such He variation is in fact needed to reproduce the observed HB morphology of the cluster.
Therefore a mass difference between the FG and the helium enriched SG populations provides a plausible and dynamically robust interpretation of our observational result. 

Our analysis has shown that small differences in the average mass may play a critical role in shaping the radial
distributions of LE-MPs in dynamically evolved systems and therefore they should be carefully accounted for in
the interpretation of the observed radial distribution differences, in particular for dynamically evolved systems. A more extended survey of simulations covering a broader range of initial conditions 
would be necessary to further explore these issues and provide a more comprehensive characterization of the extent of 
the expected differences in the spatial distribution for different radial coverage and dynamical ages. On the observational side, additional studies aimed at exploring the spatial distributions of extreme SG populations are essential to provide additional constraints for the study of the dynamics of multiple populations and the manifestations of the possible differences in the masses of FG and SG stars.

\vspace{1.0cm}
ED warmly thank Peter Stetson for kindly providing ground based photometric catalogs of M~80. 
The authors thank the anonymous referee for the careful reading of the paper and the constructive comments.
 
\software{CataXcorr (P. Montegriffo - \url{http://davide2.bo.astro.it/~paolo/Main/CataPack.html}, DAOPHOT IV \citet{stetson87}, McLuster \citet{kuepper11}, NBODY6 \citet{aarseth2003, nitadori2012}}

%% This command is needed to show the entire author+affilation list when
%% the collaboration and author truncation commands are used.  It has to
%% go at the end of the manuscript.
%\allauthors

%% Include this line if you are using the \added, \replaced, \deleted
%% commands to see a summary list of all changes at the end of the article.
%\listofchanges

\end{document}